\documentclass[fleqn,12pt,twoside]{article}
\usepackage{usehep}
\usepackage{graphicx}
\usepackage[figuresright]{rotating}

\newcommand{\AmS}{{\protect\the\textfont2
  A\kern-.1667em\lower.5ex\hbox{M}\kern-.125emS}}
\def\C{c{\bar c}}

\def\e{\epsilon}

\def\be{\begin{equation}}
\def\ee{\end{equation}}
\def\lsim{\raise0.3ex\hbox{$<$\kern-0.75em\raise-1.1ex\hbox{$\sim$}}}
\def\gsim{\raise0.3ex\hbox{$>$\kern-0.75em\raise-1.1ex\hbox{$\sim$}}}

\begin{document}

\thispagestyle{empty}

\title{~\vskip -4.5cm
~\vskip 2.2cm
{\bf Limits of Confinement:\\
\vspace*{0.05cm}
The First 15 Years of Ultra-Relativistic Heavy Ion Studies}}

\author{H.\ Satz 
\address{Fakult\"at f\"ur Physik, Universit\"at Bielefeld,\\
Postfach 100 131, D-33501 Bielefeld, Germany}%
\thanks{Opening talk at Quark Matter 2002, Nantes, France, 
July 17 - 24, 2002}}

\maketitle

\begin{abstract}

The study of high energy nuclear collisions has entered a new stage
with RHIC; it therefore seems a good time to ask what we have learned
from the experimental results obtained up to now. I recall what we had
expected to find when the SPS and AGS programs were started, summarize
what actually was found, and then try to assess what we have learned
from the results.

\end{abstract}

\section{The Beginning}

It all began with the idea of an intrinsic limit to hadron
thermodynamics. During the past fifty years, different conceptual
approaches had led to an ultimate temperature of strongly interacting
matter. Pomeranchuk~\cite{Pom} first obtained it from the finite
spatial extension of hadrons: a hadron can only have an independent
existence if it has an independent volume. Then Hagedorn~\cite{Hage}
arrived at a limiting temperature by postulating a self-similar
hadronic resonance composition: a resonance consists of resonances
which consist of resonances, and so on. The resulting excitation
spectrum was later also derived in the dual resonance
model~\cite{DRM}.  With the advent of the quark infrastructure of
hadrons and of quantum chromodynamics, it became clear that the
ultimate temperature found in all these considerations was really a
transition point to a new state of matter, to a plasma of deconfined
quarks and gluons~\cite{C-P}.

Statistical QCD, in particular in the finite temperature lattice
formulation, has subsequently confirmed this hypothesis: at a now
rather well determined temperature (for vanishing baryon density, $T_c
\simeq 150-180$~MeV), strongly interacting matter undergoes a
transition from a medium of color singlet hadronic constituents to one
of deconfined colored quarks and gluons~\cite{Karsch}. The energy
density at the transition point was found to be $\e(T_c) \simeq
0.7-1.0$~GeV/fm$^3$.  Moreover, the transition turns a hadronic state
of spontaneously broken chiral symmetry into a quark-gluon plasma in
which this symmetry is restored: at $T_c$, the effective constituent
quark mass of some 0.3~GeV vanishes, and we recover the bare quark
mass of the QCD Lagrangian.

The obvious desire to test this fascinating phase structure of
strongly interacting matter first led to the fixed target experiments
at the AGS in Brookhaven (with $\sqrt s \simeq 5$~GeV) and at the
CERN-SPS (with $\sqrt s \simeq 20$~GeV). In 1986/87, light ion beams
on heavy ion targets started the program, and in 1994/95, heavy ion
beams followed.  Today, much of this program is concluded. So, what
have we learned during the past fifteen years? In this opening talk, I
will address that question by asking:
\begin{itemize}
\parskip=0.pt \parsep=0.pt \itemsep=0.pt
\item What did we expect to find?
\item What did we find?
\item What does that tell us?
\end{itemize}
\noindent
In my report, I will first recall briefly the expectations concerning
signatures at the beginning of the experimental heavy ion program at
the AGS and SPS in 1986 and then summarize what had really been found
when it was (in first generation experiments) completed in 2000.
Following this, I will try to indicate what conclusions can be drawn
from these results, for the conditions reached, from the hard probes
of the early stages and from the observed hadronisation pattern at
freeze-out.

\section{Hopes}

\begin{figure}[htb]
\vglue-4mm
\centering
\resizebox{0.50\textwidth}{!}{%
\includegraphics*{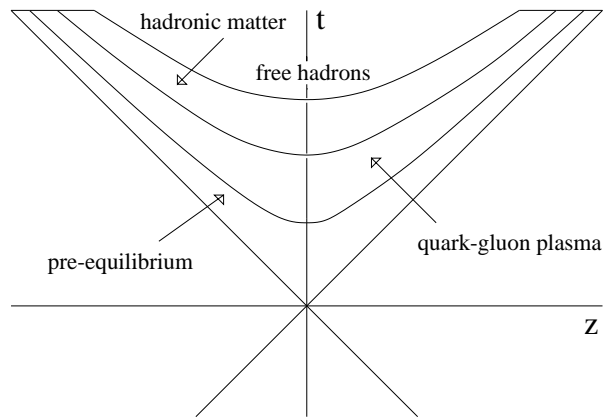}}
\vglue-8mm
\caption{The expected evolution of a nuclear collision}
\label{evo}
\vglue-4mm
\end{figure}

The evolution of a high energy nucleus-nucleus collision was pictured
in the form shown in Fig.~\ref{evo}. After a rather short
equilibration time $\tau_0 \simeq 1$~fm, the presence of a thermalized
medium was assumed, and for sufficiently high initial energy
densities, this medium would be in the quark-gluon plasma phase.

The initial energy density of the produced medium at the time of 
thermalization was (and still is) generally determined by the Bjorken
estimate~\cite{Bjorken83}
\be
\e = \left({dN_h \over dy}\right)_{y=0} {w_h \over \pi R_A^2 \tau_0},
\label{2.1}
\ee 
where $(dN_h/dy)_{y=0}$ specifies the number of hadronic secondaries
emitted per unit rapidity at mid-rapidity and $w_h$ their average
energy. The effective initial volume is determined in the transverse
plane by the nuclear radius $R_A$, and longitudinally by the formation
time $\tau_0$ of the thermal medium.

The temperature of the produced medium was assumed to be observable
through the transverse mass spectrum of thermal dileptons and the
momentum spectrum of thermal photons~\cite{Shuryak,Keijo}. The
observation of thermal dilepton/photon spectra would also indicate
that the medium was indeed in thermal equilibrium. The functional form
of the spectra is the same for radiation from hadronic matter and from
a QGP; but the observed rates and temperatures were expected to differ
in the two cases. It was clear from the beginning that these signals
would be emitted during the entire thermal evolution of the system,
making a separation of different phases of origin not very
straight-forward.

The determination of the nature of the hot initial phase required a
signature sensitive to deconfinement. It was argued that in a
deconfined medium the J/$\psi$ would melt through color
screening~\cite{MS} and that, therefore, QGP production should lead to
a suppression of J/$\psi$ production in nuclear collisions, compared
to the rates extrapolated from $pp$ data.  Similarly, the QGP was
expected to result in a higher energy loss for a fast passing color
charge than a hadronic medium, so that jet quenching~\cite{jets}
should also signal deconfinement.

The behavior of sufficiently short-lived resonances, in particular the
dilepton decay of the $\rho$, was considered as a viable tool to study
the hadronic medium in its interacting stage and thus provide
information on the approach to chiral symmetry
restoration~\cite{rho-chiral}.

The expansion of the hot medium was thought to be measurable through
broadening and azimuthal anisotropies of hadronic transverse momentum
spectra~\cite{flow}. The size and age of the source at freeze-out was
assumed to be obtainable through Hanbury-Brown--Twiss (HBT)
interferometry based on two-particle correlations~\cite{HBT}.  It was
expected that increasing the collision energy would increase the
density and hence the expansion of the produced medium, so that the
HBT radii should grow with $\sqrt s$.

The final interacting hadronic medium was discussed in terms of an
ideal resonance gas, following an old suggestion~\cite{B-U} brought to
hadron physics by Hagedorn~\cite{Hage}: an interacting system of
elementary constituents can be replaced by a non-interacting gas of
resonances, provided the elementary interactions are
resonance-dominated. This would provide the relative abundances of all
hadron species in terms of just two parameters, the temperature and
the baryon number density. One particularly interesting feature here
was the fact that in elementary hadronic interactions, an overall
reduction of strangeness production was observed. Nuclear collisions,
in particular if leading to a QGP as initial stage~\cite{Rafelski},
were expected to remove this reduction and lead to strangenes
production in accord with thermal predictions.

\section{Facts}

\begin{figure}[ht!]
\centering
\resizebox{0.5\textwidth}{!}{%
\includegraphics*{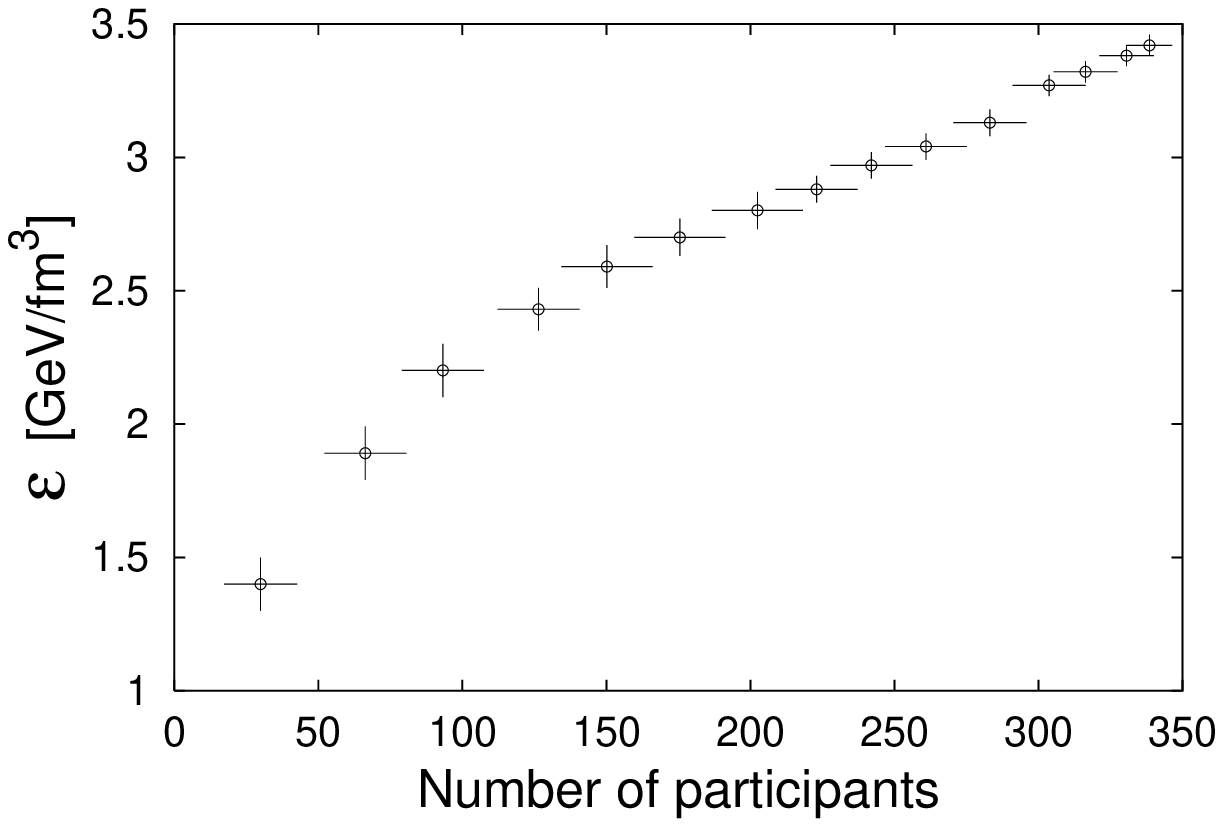}}
\vglue-8mm
\caption{The energy density in Pb-Pb collisions at
$\sqrt{s}=17$~GeV~\cite{NA50,Nardi}.}
\label{e-dens}
\vglue6mm
\centering
\begin{tabular}{cc}
\resizebox{0.48\textwidth}{!}{%
\includegraphics*{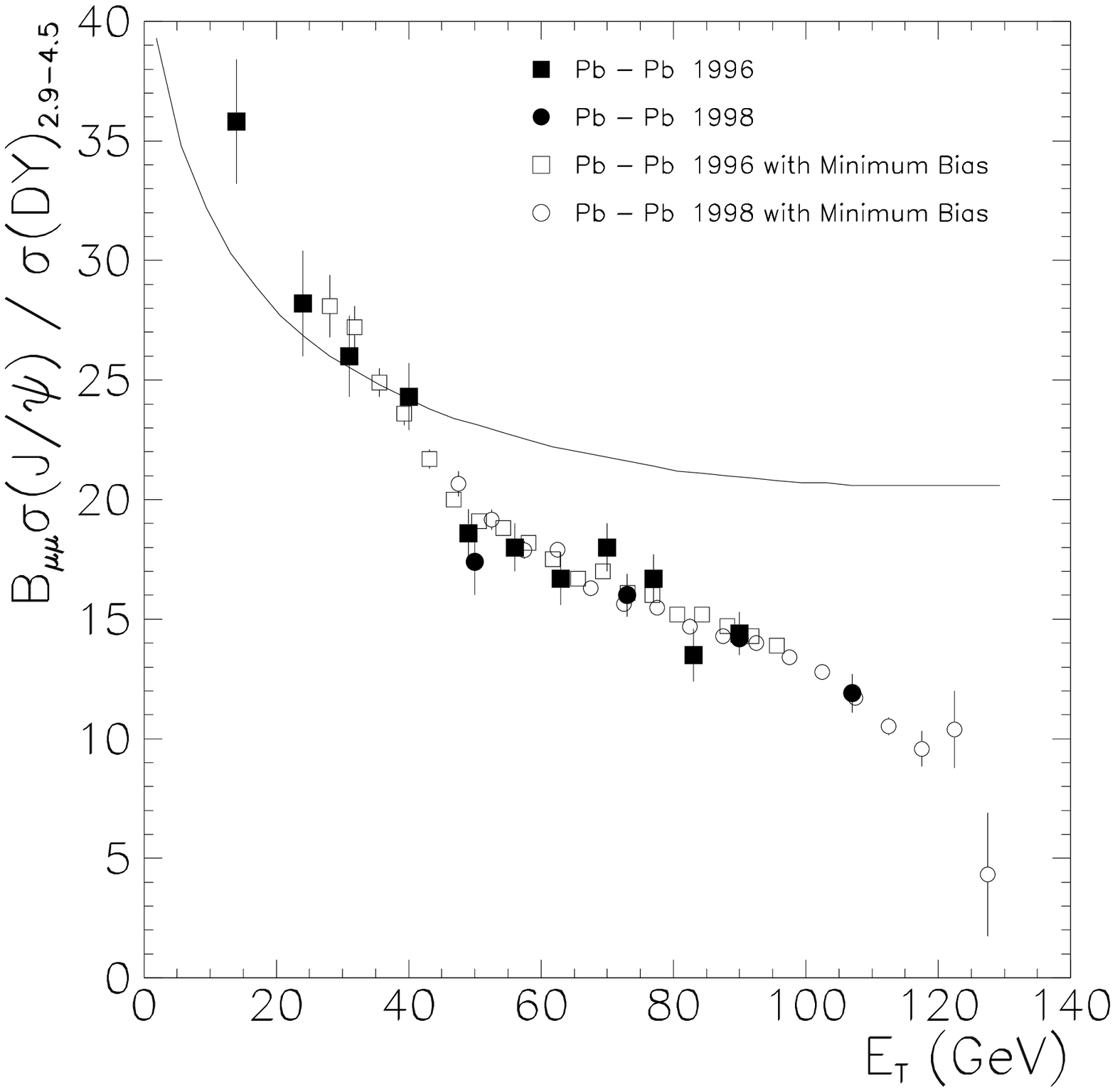}}
&
\resizebox{0.48\textwidth}{!}{%
\includegraphics*{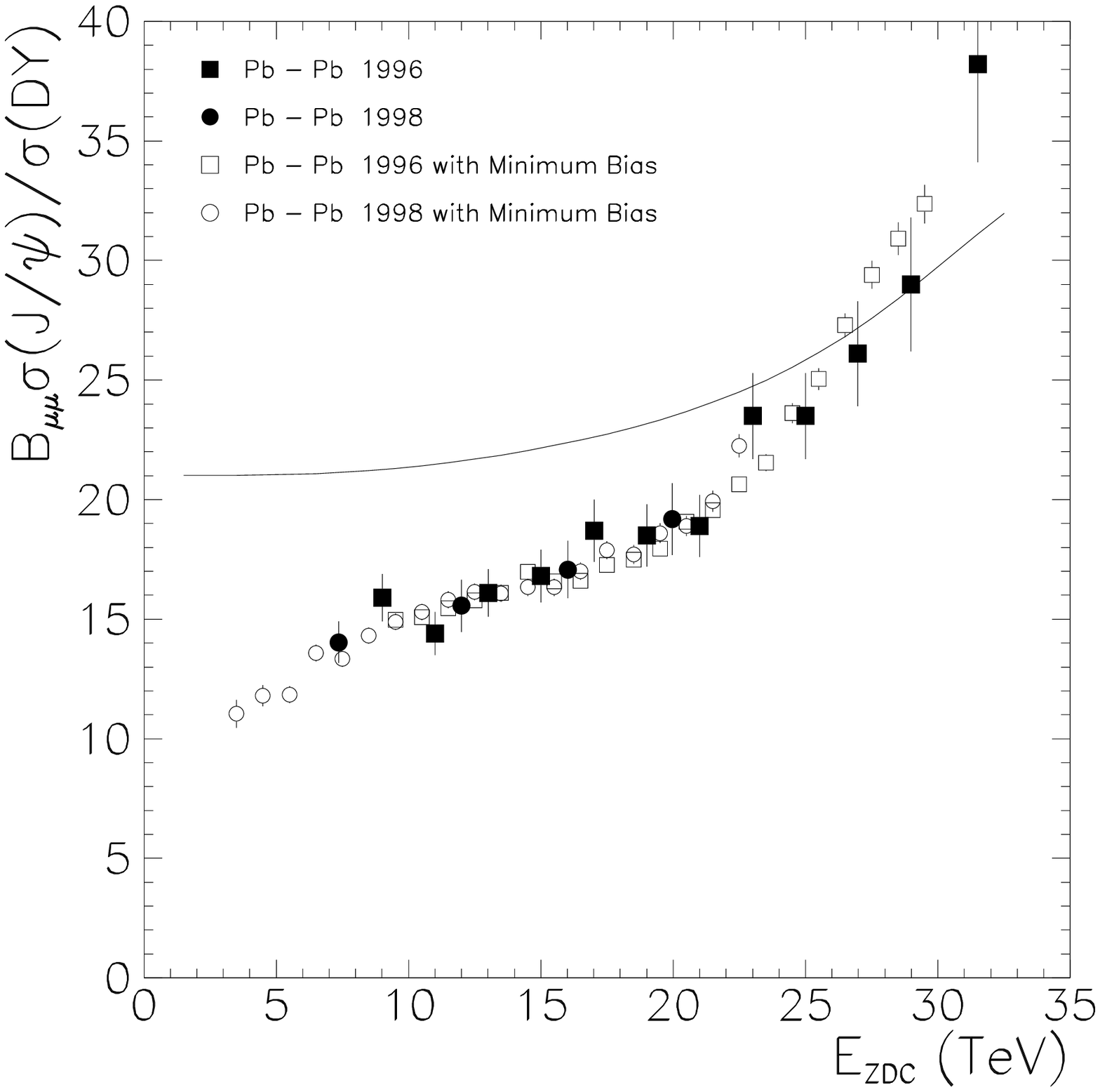}}
\end{tabular}
\vglue-8mm
\caption{The ratio of J/$\psi$ to Drell-Yan production, measured by
NA50, in Pb-Pb collisions at $\sqrt s = 17$ GeV, vs.\ $E_T$ (left) and
$E_{ZDC}$ (right).  The solid line indicates the extrapolation of the
normal nuclear absorption infered from p-A collisions~\cite{NA50}.}
\label{J/psi}
\vglue-8mm
\end{figure}

The initial energy density, as specified by the Bjorken estimate,
Eq.\ (\ref{2.1}), was measured in almost all SPS experiments. In
Fig.~\ref{e-dens} we show $\e$ as function of centrality, determined
by the number of participant nucleons~\cite{NA50,Nardi}; it covers the
range from somewhat above 1 to almost 3.5~GeV/fm$^3$. Finite
temperature lattice calculations, as already mentioned, give for the
energy density at deconfinement, $\e(T_c)$, values around or slightly
below 1~GeV/fm$^3$~\cite{Karsch}. However, also high energy $p\bar p$
collisions lead to energy density estimates well above 1~GeV/fm$^3$.

Thermal photon or dilepton spectra have so far not been unambiguously
identified. Some photon excess over the expected hadronic decay yield
has been reported~\cite{photondata}, but its interpretation is still
open.  An observed excess of dileptons in the mass range between the
$\phi$ and the J/$\psi$~\cite{hm-dileptons} has been attributed to
thermal emission during the evolution of the system~\cite{R-S}, but
only a small fraction would be due to the hot early stage; also here, 
more data is needed to understand the origin of the observed effects.

\begin{figure}[ht!]
\vglue-4mm
\centering
\begin{minipage}[t]{0.48\textwidth}
\centering
\resizebox{0.98\textwidth}{!}{%
\includegraphics*{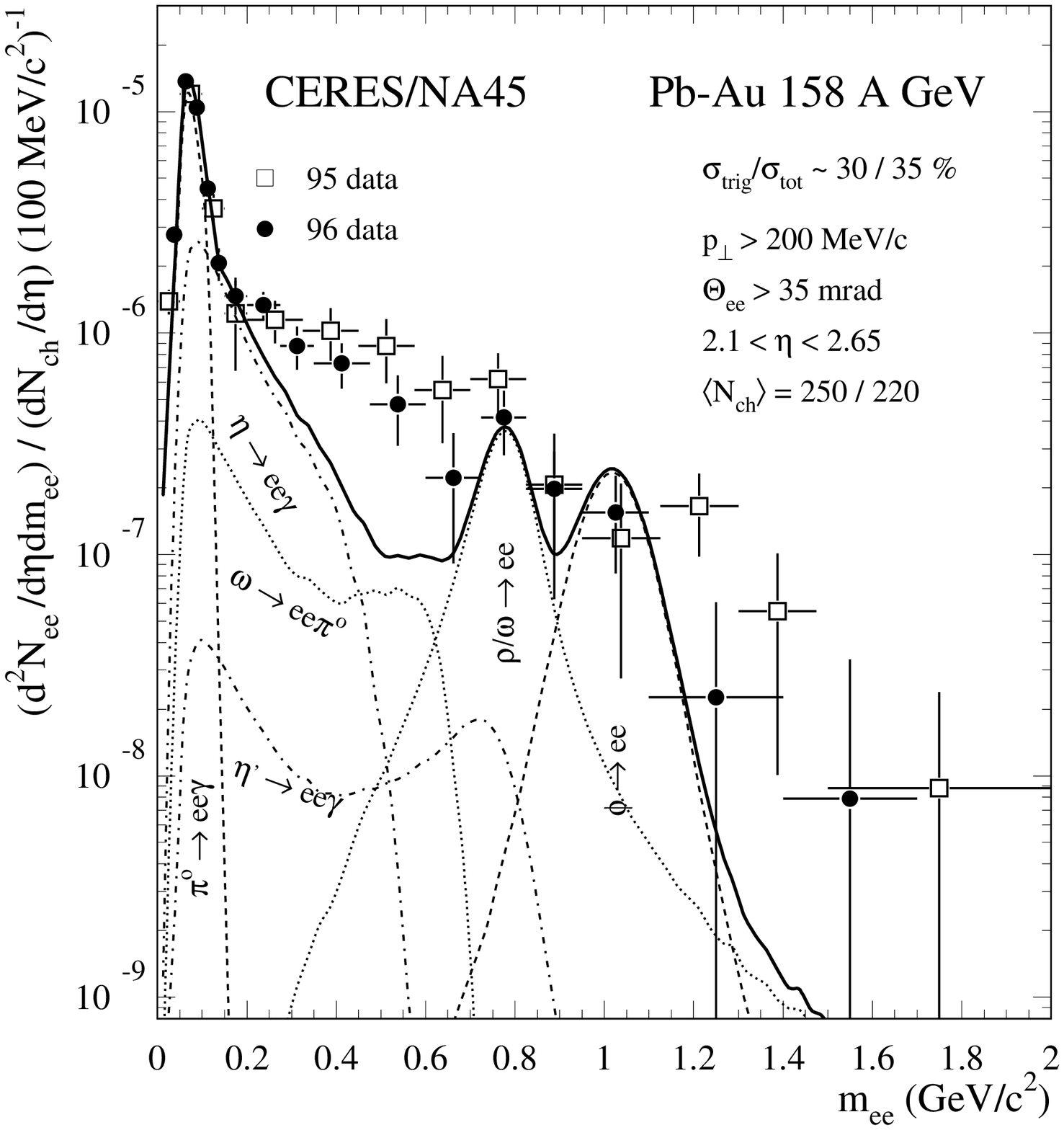}}
\vglue-8mm
\caption{The dilepton spectrum in Pb-Au collisions at $\sqrt s =
17$~GeV, compared to the expected yield (solid line) from known
hadronic sources~\cite{Ceres}.}
\label{rho}
\end{minipage}
\hfill
\begin{minipage}[t]{0.48\textwidth}
\centering
\resizebox{0.88\textwidth}{!}{%
\rotatebox{90}{%
\includegraphics*{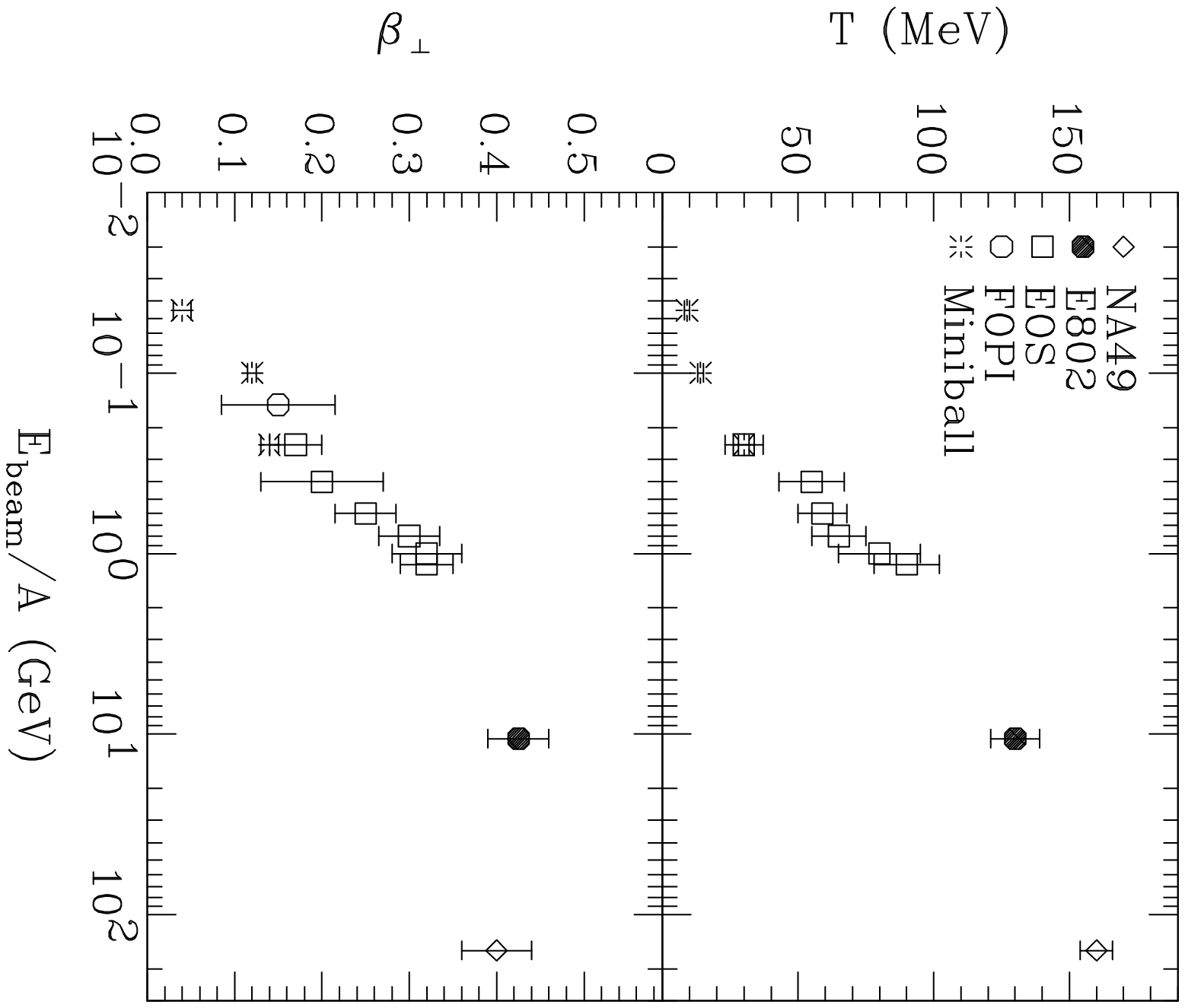}}}
\vglue-8mm
\caption{Hadronization temperature, $T$, and transverse flow measure,
$\beta_T$, as function of the beam energy~\cite{Pavel}.}
\label{flow}
\end{minipage}
\vglue-4mm
\end{figure}

J/$\psi$ production was found to be suppressed in O-U, S-U and then
Pb-Pb collisions; the suppression always increases with
centrality~\cite{NA38/50}.  Studying p-A collisions, it was observed
that already normal nuclear matter leads to reduced charmonium
production. Extrapolating this `normal' suppression to $AB$
interactions is enough to account for the observed yields up to
central S-U collisions. In central Pb-Pb collisions, however, an
additional `anomalous' suppression was observed~\cite{NA50}.
Peripheral Pb-Pb collisions follow the normal pattern; then, with
increasing centrality (measured either through the associated
transverse energy or through the number of participating nucleons),
there is a pronounced onset of further suppression
(Fig.~\ref{J/psi}). The very peripheral data points shown in this
figure are somewhat above the normal suppression curve; this was shown
to be due to beam-air interactions and is no longer the case in the
data collected in year 2000, when the target was placed in
vacuum~\cite{Ramello}.

The dilepton mass spectrum in the region below the $\rho$ peak was
indeed found to differ considerably from the yield and form expected
from known hadronic sources~\cite{Ceres}, indicating the presence of
in-medium resonance modifications (Fig.~\ref{rho}). This `low mass
dilepton enhancement' is observed in S-U and Pb-Pb collisions, and for
the latter at beam energies of 40~GeV as well as of 160~GeV.
The broadening of transverse momentum spectra, expected as consequence
of transverse flow, was observed in the predicted form of increasing
broadening with hadron mass. If parametrized in terms of radial
flow~\cite{Pavel}, the extracted flow velocity $\beta_T$ appears to
saturate with increasing collision energy (Fig.~\ref{flow}).

\begin{figure}[ht!]
\vglue-4mm
\centering
\begin{minipage}[t]{0.48\textwidth}
\centering
\resizebox{0.95\textwidth}{!}{%
\includegraphics*{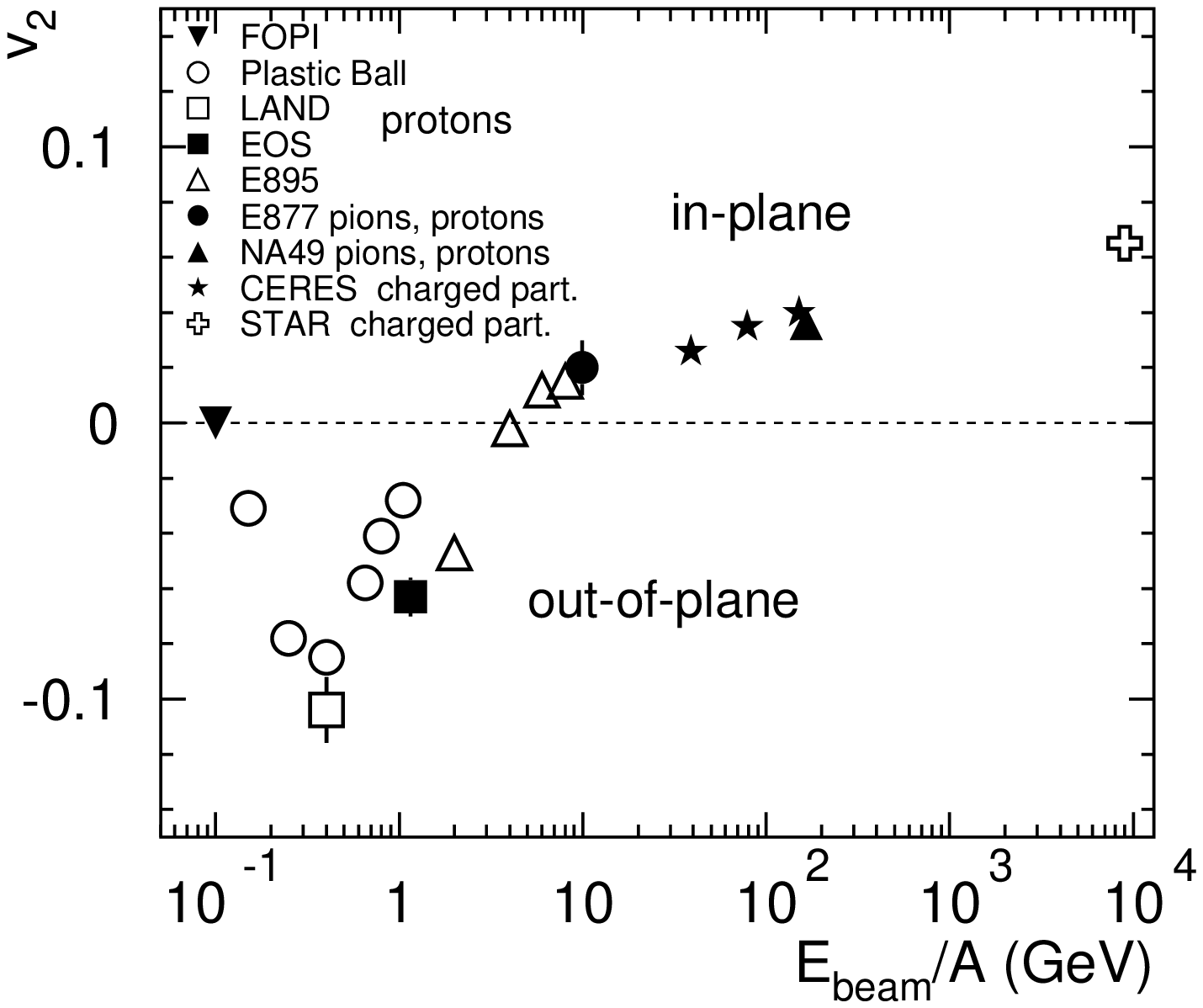}}
\vglue-8mm
\caption{Elliptic flow measure, $v_2$, as function of the beam
energy~\cite{Appel}.}
\label{ellip}
\end{minipage}
\hfill
\begin{minipage}[t]{0.48\textwidth}
\centering
\resizebox{0.8\textwidth}{!}{%
\includegraphics*{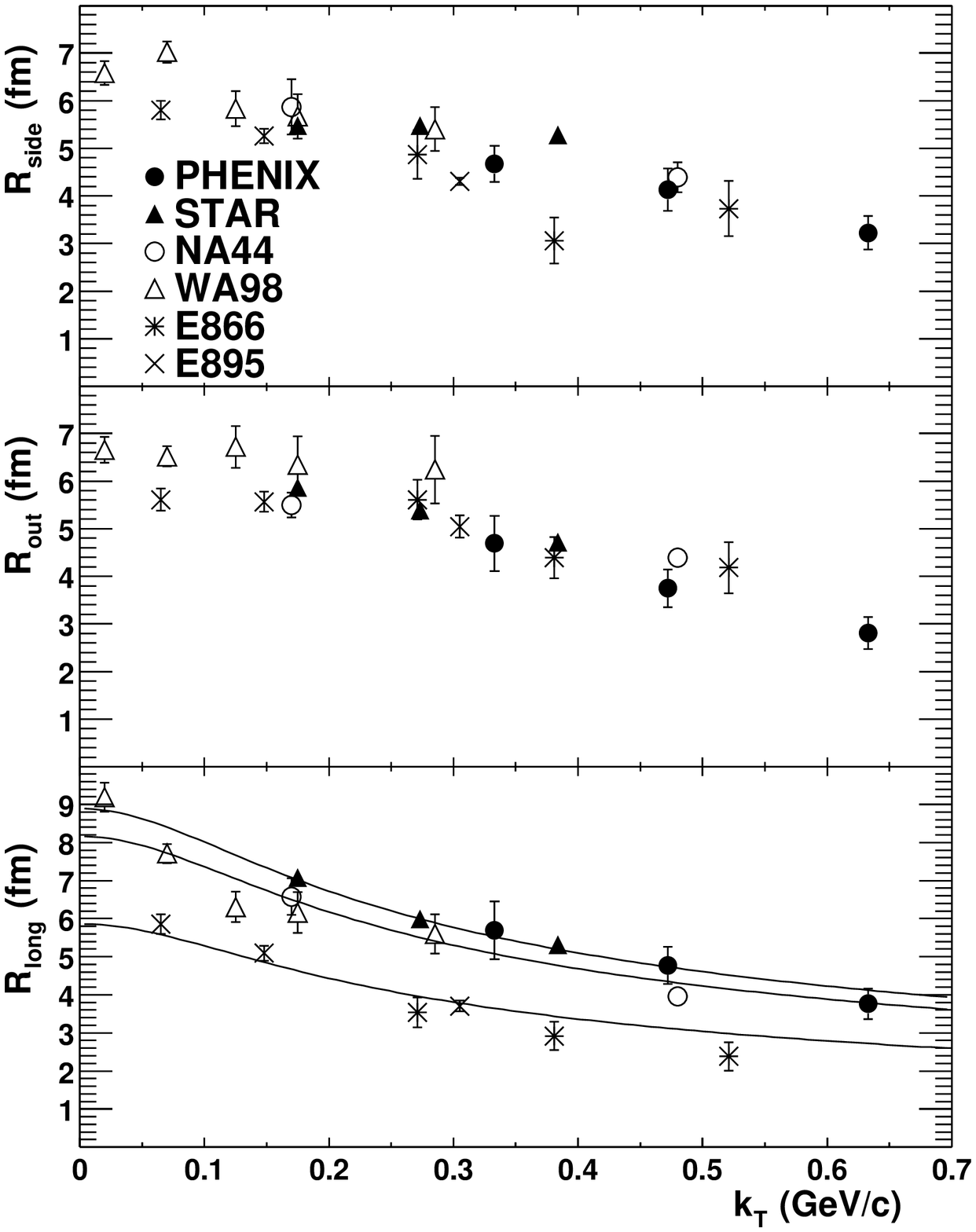}}
\vglue-8mm
\caption{HBT radii at different collision energies~\cite{adcox}.}
\label{HBT}
\end{minipage}
\vglue4mm
\centering
\resizebox{0.8\textwidth}{!}{%
\rotatebox{270}{%
\includegraphics*{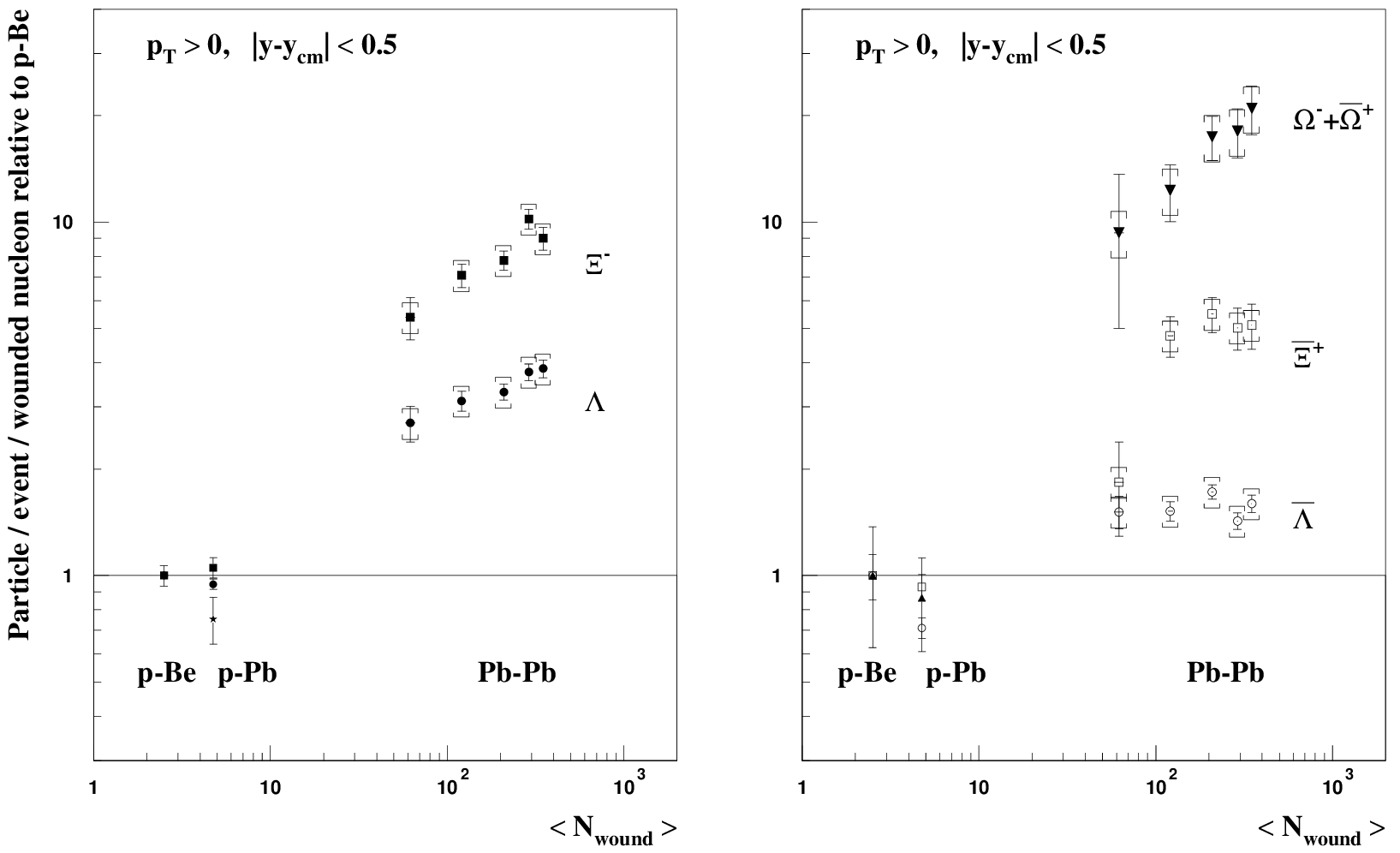}}}
\vglue-8mm
\caption{Strange hyperon production per participant nucleon, normalized
to the ratio from p-Be, as function of the number of participant nucleons, 
as measured by NA57~\cite{antinori}.} 
\label{Antinori}
\vglue-4mm
\end{figure}

The transverse momentum spectra, moreover, also showed the azimuthal
anisotropy predicted for non-central collisions. The behavior shown in
Fig.~\ref{ellip} indicates at low collision energy a reduced emission
in the transverse direction, where spectator nucleons are present; at
high energy, there is enhanced production in the direction of the
higher pressure gradient as determined by the anisotropic interaction
volume~\cite{Appel}.

The expected increase of source size in HBT correlation studies has
not been observed; instead, one finds that, with increasing collision
energy, the observed source radii are essentially determined by those
of the involved nuclei~\cite{adcox}.  Thus one finds $R_{\rm side}
\simeq R_{\rm out} \simeq 5-6$~fm for Au-Au/Pb-Pb collisions from AGS
to SPS (and on to RHIC), as seen in Fig.~\ref{HBT}.  This is to be
contrasted to predicted values for Pb-Pb collisions ranging from 14~fm
for the SPS to 20~fm for RHIC, if freeze-out occurs when the mean free
path of a hadron exceeds the system size~\cite{stock}
\be
R_f \simeq 0.7
\left({dN_h \over dy} \right)^{1/2}.
\label{2.2}
\ee
For a freeze-out at the ideal pion gas energy density, one has $R_f
\sim (dN_h /dy)^{1/3}$, which increases from 9 to 12~fm between SPS
and RHIC energies. Another unexpected feature is that the outward and
sideward radii, $R_{\rm out}$ and $R_{\rm side}$, whose difference
could provide information about the life-time of the emitting medium,
appear to be approximately equal. The HBT source radii in nuclear
collisions thus seem to be energy-independent and fixed by the initial
nuclear size~\cite{Adamova}.

\begin{figure}[ht!]
\vglue-4mm
\centering
\begin{minipage}[t]{0.48\textwidth}
\centering
\resizebox{0.95\textwidth}{!}{%
\includegraphics*{wrob.eps}}
\vglue-8mm
\caption{Wroblewski strangeness measure for different collision
configurations~\cite{BC}.}
\label{Wrob}
\end{minipage}
\hfill
\begin{minipage}[t]{0.48\textwidth}
\centering
\resizebox{0.95\textwidth}{!}{%
\includegraphics*{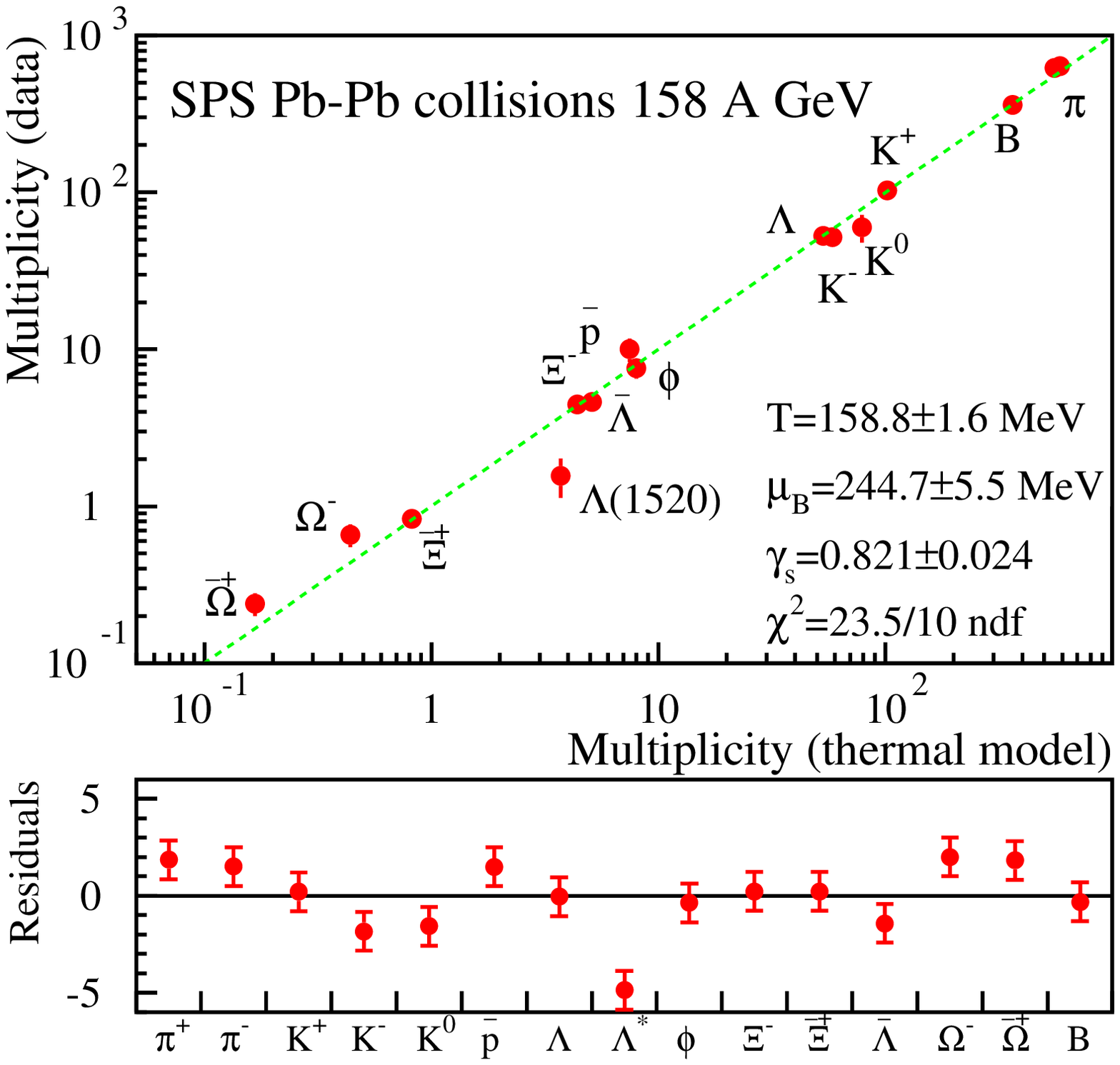}}
\vglue-8mm
\caption{Thermal hadronization and species abundances, from
Ref.~\cite{becattini}.}
\label{becat}
\end{minipage}
\vglue8mm
\centering
\resizebox{0.57\textwidth}{!}{%
\includegraphics*{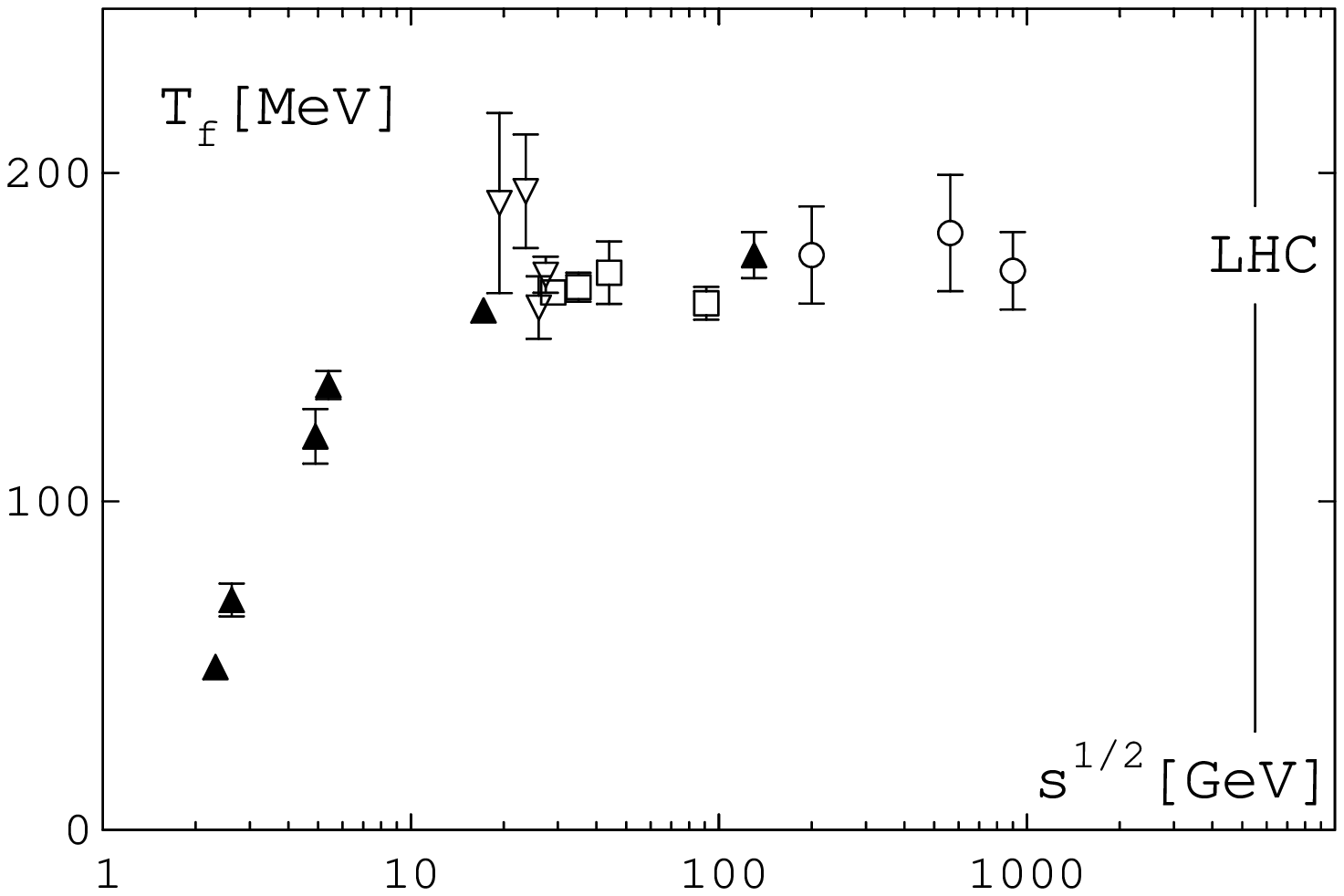}}
\vglue-8mm
\caption{Hadronization temperature as function of c.m.s.\ energy, for
$e^+e^-$ (squares), $pp$ (open triangles), $p\bar p$ (circles) and
$AA$ collisions (filled triangles)~\cite{ROP}.}
\label{Hage}
\vglue-4mm
\end{figure}

Hadron abundances showed the expected enhancement of strangeness
production; in Fig.~\ref{Antinori} we show the most striking example,
where the production of strange baryons is increased up to 10 times
and more in comparison to $pp$ rates~\cite{antinori}. In
Fig.~\ref{Wrob}, it is seen that, quite generally, the Wroblewski
measure for strangeness content is about a factor two higher in $AA$
collisions than in elementary particle reactions. Of course, the role
of the changing baryon density has to be kept in mind when comparing
strangeness production in $pp$ and $AA$ collisions, or among $AA$
collisions of different energies.

Nevertheless, in all cases the abundance of the different hadron
species is very well described by an ideal resonance gas. A fit to the
latest NA49 data is shown in Fig.~\ref{becat}~\cite{becattini}.  The
hadronization temperature increases with increasing collision energy
and converges for all hadron production, from $e^+e^-$ and $pp/p \bar
p$ to high energy $AA$ collisions, to a stable value around
150--180~MeV; see Figs.~\ref{flow} and~\ref{Hage}. The associated
baryo-chemical potential decreases when going from AGS to SPS and on
to RHIC.

\section{Lessons}

What can we learn from the results just listed? On the one hand, this
is still a rather subjective issue, and so I proceed with due
apologies to all those who feel that I have misunderstood what nature
is trying to tell us. On the other hand, our theoretical understanding
is presently at what might be a crucial point. It was and is generally
thought that high energy nuclear collisions will produce a thermalized
medium of deconfined quarks and gluons, so that we can test the
predictions of finite temperature statistical QCD. In recent years,
however, the role of the primary parton state has been emphasized in 
two independent but closely related approaches, color glass
condensation~\cite{CGC,Larry} and parton percolation~\cite{Torino,DFPS}. 
We are thus today confronted with the
interesting question of knowing what part of the observable features
of nuclear collisions are determined by the primary initial
conditions, before any thermalization may take place.

Assuming thermal equilibration, with $\tau_0 \simeq 1$ fm, the initial
energy densities in central Pb-Pb collisions at the SPS are estimated
to reach 3~GeV/fm$^3$ or more. What kind of medium can we have at this
stage --- could it be hadronic matter, a system of hadronic comovers
of some kind? I believe that the answer is a clear NO. The average
energy per hadron in these reactions is around 0.5~GeV, so that the
mentioned energy density estimates would imply a hadron density of
some 6~hadrons/fm$^3$. The hadronic mean free path in such a medium,
$\lambda = 1/n_h \sigma_h$, reaches something like $\lambda\simeq
0.06$~fm, with a typical hadron-hadron interaction cross section of
around 30~mb.  The time needed by a hadron to react to a single
collision is around 1~fm, and in that time it travels a distance $\sim
1$~fm.  In other words, the hadron undergoes almost twenty collisions
in the time it needs to react to a single one. This rules out any
medium consisting of independent hadrons~\cite{Pom,LPM}.

If the produced medium is not hadronic, does that mean that it is a
quark-gluon plasma? The answer is not as easy as we once thought, so
let us approach this problem slowly and carefully.

\begin{figure}[ht!]
\vglue-8mm
\centering
\begin{minipage}[t]{0.5\textwidth}
\centering
\resizebox{0.95\textwidth}{!}{%
\includegraphics*{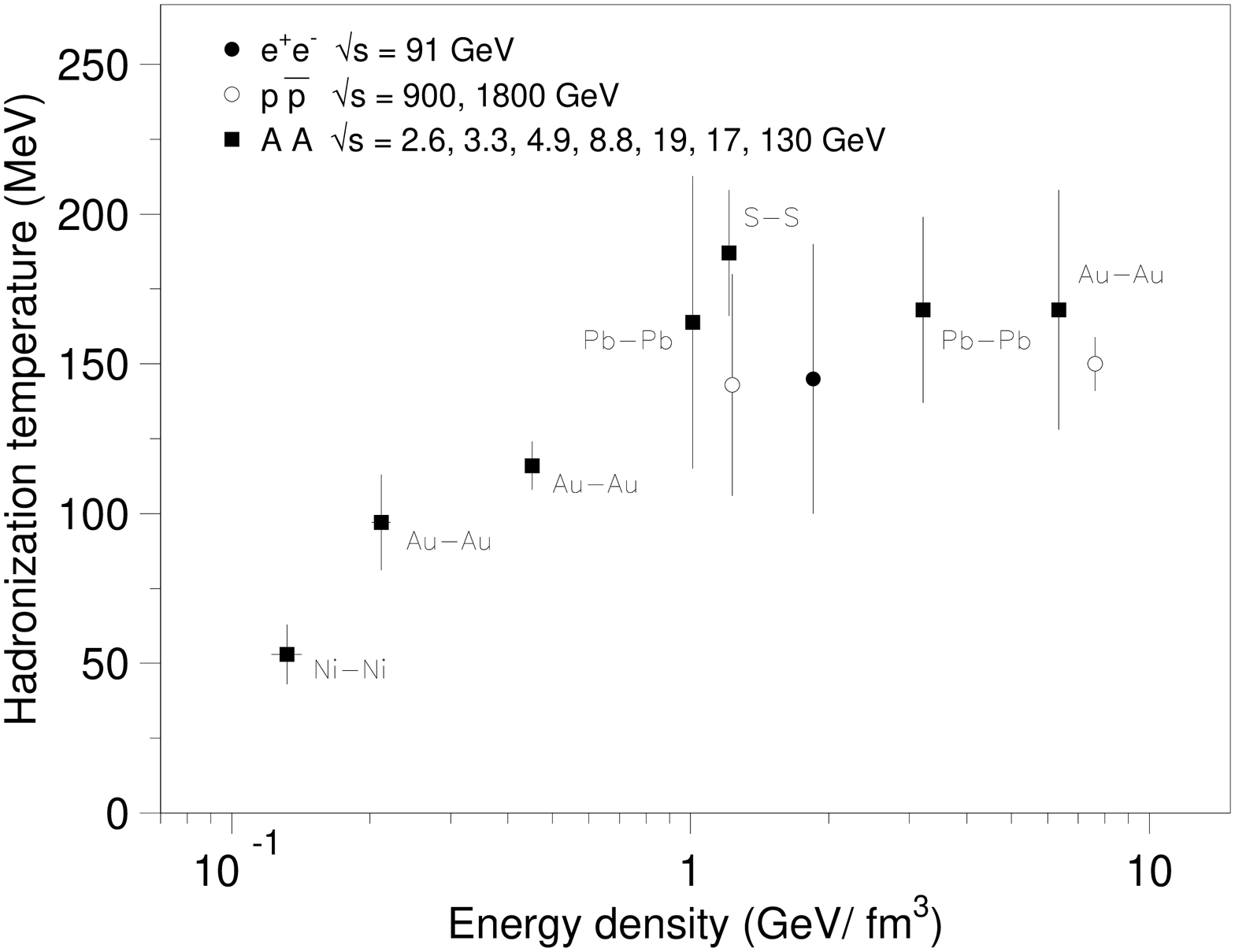}}
\vglue-8mm
\caption{Hadronization temperature vs.\ the initial energy
density, from Ref.~\cite{Sonja}.}
\label{kabana}
\end{minipage}
\hfill
\begin{minipage}[t]{0.45\textwidth}
\centering
\vglue-5.9cm
\resizebox{0.84\textwidth}{!}{%
\rotatebox{270}{%
\includegraphics*{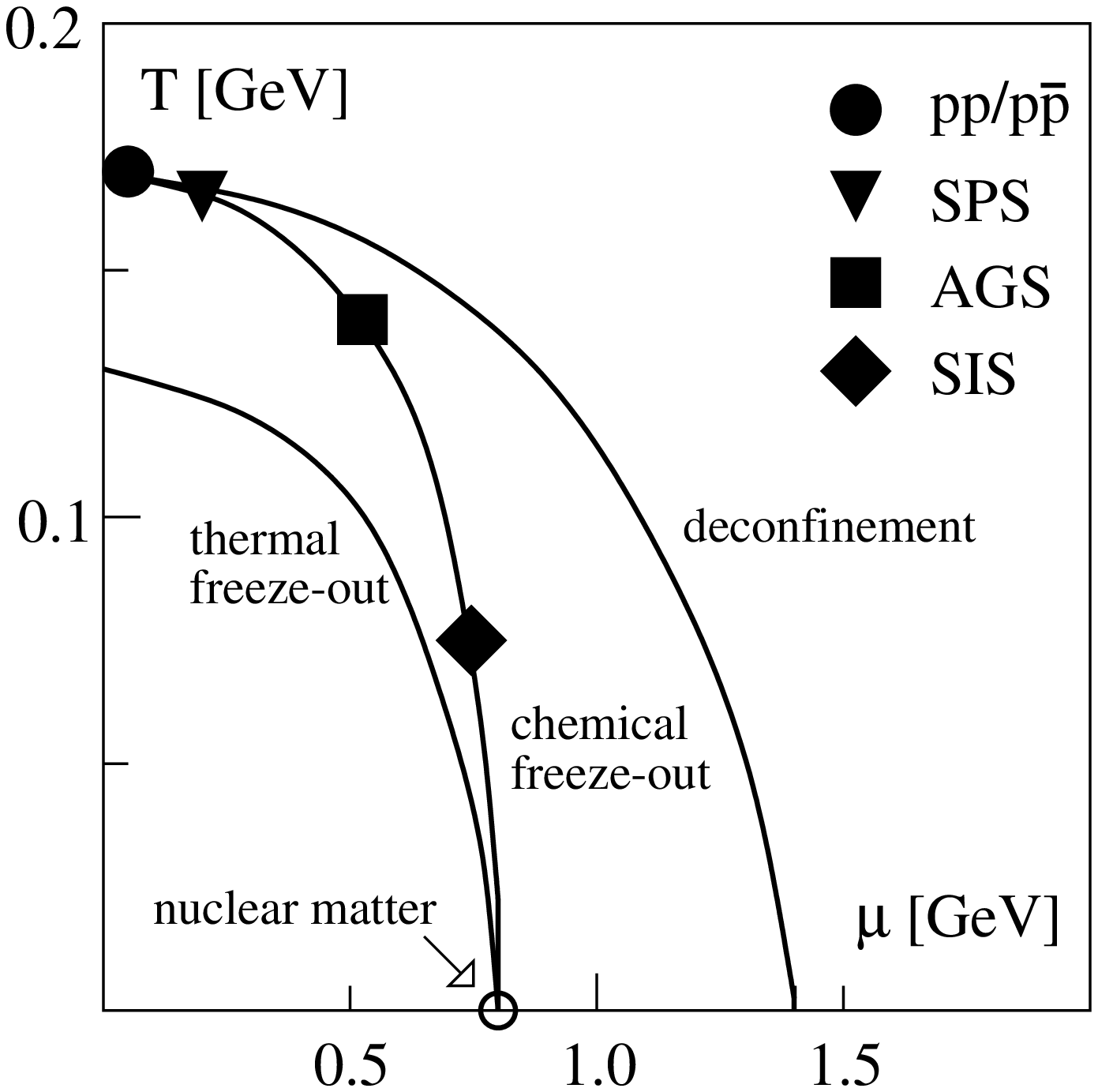}}}
\vglue-6.5mm
\caption{Deconfinement, chemical and thermal freeze-out 
vs.\ $T$ and $\mu$~\cite{stachel,redlich}.}
\label{T-mu}
\end{minipage}
\vglue-4mm
\end{figure}

In a comparative study of the hadronization temperature $T_H$ as
function of the initial energy density for different production
processes, from $e^+e^-$ and $pp/p \bar p$ reactions up to the
heaviest $AA$ collisions, we have to take into account the different
values of the associated baryo-chemical potential. Since in all cases
hadronization can be described in terms of an ideal resonance gas, we
can extrapolate the baryon-rich low energy $AA$ results
at constant entropy to $\mu_B=0$; the resulting behavior is shown in
Fig.~\ref{kabana}~\cite{Sonja}.  It is quite striking: above some
threshold value $\e_c \simeq 1$~GeV/fm$^3$, putting more energy into
the system has no effect on the critical temperature $T_H$ which
determines the hadron abundances.  In the end, Hagedorn was
right~\cite{Hage}. This is like observing that water vapor always
condenses to water at 100\,$^{\circ}$C, whatever was the initial 
temperature of the vapor.  While this shows that the water at
the condensation point gives no information about the earlier vapor
temperatures, it also indicates that the condensation process exhibits
critical behavior. In the same way, a constant $T_H$ for hadronic
matter of arbitrarily high energy densities appears to signal critical
behavior~\cite{C-P}. With this in mind, we now look once more at the
experimental numbers. The saturation temperature is 150--180~MeV, and
is reached for $\e \simeq 1$~GeV/fm$^3$; both values are in
excellent agreement with the mentioned results from lattice QCD
studies~\cite{Karsch}. For very small or vanishing baryon density,
experiment and theory thus fully agree on the boundary of hadron
physics.

Extending the resonance gas analysis of species abundances to finite
baryon density, one obtains the `chemical' freeze-out 
curve \cite{PBM-f,C-R}
in the $T-\mu_B$ plane; it is illustrated schematically in
Fig.~\ref{T-mu}. Also included in this figure are the thermal
freeze-out curve (obtained from a flow analysis of transverse momentum
spectra) and the expected deconfinement transition
curve~\cite{stachel}.  Some first and quite promising attempts of
lattice gauge theory to study deconfinement at non-zero baryo-chemical
potential, $\mu_B$, agree with the deconfinement curve for $\mu_B \leq
0.8$~GeV~\cite{Fodor,Allton}. It is clear that the deconfinement curve
must lie above the freeze-out curve at large $\mu_B$: the interactions
in cold dense nuclear matter are dominated by baryon repulsion and
hence cannot be described in terms of an ideal resonance gas. At
$T\!=\!0$, freeze-out occurs at normal nuclear matter density, where
baryon repulsion and attractive (pion exchange) interactions just
balance each other. Neutron stars are examples of confined states of
higher baryon densities.

Since Figs.~\ref{Hage}, \ref{kabana} and~\ref{T-mu} include both data
from elementary particle and from nuclear collisions, statistical
hadron abundances alone are not sufficient to indicate the presence of
a large-scale system. Therefore, it is necessary to check if $AA$
collisions produce large thermal systems or just the sum of the
smaller ones. The answer to that question is obtained through the
study of strangeness production. In elementary particle collisions,
there is an overall suppression of strange particle production
compared to the predictions of the grand canonical form of an ideal
resonance gas.  This can be interpreted as the consequence of local
strangeness conservation: if only one $s \bar s$ pair is formed in the
gas, the corresponding Boltzmann factor should be $\exp\{-2m_s/T\}$,
instead of the grand canonical form $\exp\{-m_s/T\}$. If nuclear
collisions lead to a connected large-scale system, strangeness
conservation can occur between the products from different nuclei,
leading to the grand canonical Boltzmann factor.  As a consequence, we
expect for the Wroblewski measure in elementary particle collisions
\be
\lambda_s \equiv {2 <s \bar s> \over <u \bar u > + <d \bar d>} 
\simeq \exp\{-2m_s/T\} \simeq 0.2,
\label{wrob1} 
\ee
while central high energy $AA$ collisions should approach 
\be
\lambda_s \simeq \exp\{-m_s/T\} \simeq 0.4.
\label{wrob2}
\ee
Keeping in mind that the different baryon densities also lead to
modifications, we see in Fig.~\ref{Wrob} that Eqs.~(\ref{wrob1})
and~(\ref{wrob2}) are in fact quite well satisfied. We thus conclude
that nuclear collisions indeed produce large-scale connected systems.

The observed behavior of the hadronic transverse momentum spectra is
also often taken as indication for collective behavior of the produced
medium. In particular, azimuthal asymmetries seem to indicate that
global effects influence the spatial hadron emission: in non-central
collisions, there is more (elliptic) flow in the direction of the
greater pressure gradient. For the time being, however, it is not so
easy to reconcile a constant radial flow velocity at high energy
(Fig.~\ref{flow}) and energy-independent HBT radii of nuclear size
(Fig.~\ref{HBT}) with an initial state of partonic matter at higher
and higher energy densities. Apparently a change of conditions in the
pre-hadronic phase, whatever its nature, is not reflected in radial
flow or source size.

We conclude that there seems to be a universal limit to the
hadronization regime in high energy nuclear collisions; the abundances
of the produced hadron species, including the strange ones, are
determined by values of temperature and energy density in good
agreement with what is known about the confined region of the QCD
phase diagram. As function of the incident energy, the hadronization
temperature $T_H$, the strangeness content $\lambda_s$ and the
transverse momentum broadening (radial flow velocity $\beta_T$) all
seem to converge to constant values.  The value of the temperature
$T_H \simeq 150 - 180$~MeV and of the corresponding energy densities
agree with the confinement/deconfinement transition values obtained in
lattice QCD. The strangeness content agrees with what is expected from
the value of the Boltzmann factor $\exp\{-m_s/T\}$ at $T=T_H$. In
fact, even the low mass dilepton enhancement appears to remain
essentially invariant under changes of nuclear size and of collision
energy (in the SPS range). The origin of the effect is not yet really
clear (see, however, \cite{B-W}), but low mass dilepton production 
does seem to behave in the
same way as the mentioned hadronic observables.  Since it may be the
most sensitive to changes in the initial baryon
density~\cite{Wambach}, this observation could be a reflection of the
weak $\mu$-dependence of the confinement limit in the small $\mu$
region.

Before analyzing the earlier stages of the nuclear collisions, we
briefly mention attempts to include charmonium production in a
statistical hadronization scheme, either by assuming thermal formation
at the freeze-out point $T=T_H$ of the resonance gas~\cite{GG}, or by
assuming a thermal distribution of the produced $\C$ pairs among the
different possible open and hidden charm channels~\cite{PBM-S}. To
evaluate the viability of these approaches, we recall that J/$\psi$
production from $pp$ to central S-U and peripheral Pb-Pb collisions is
seen to be in accord with the normal nuclear suppression determined in
p-A interactions. In contrast, more central Pb-Pb collisions show a
well-defined onset of anomalous suppression in a narrow centrality
range, seen both in the ratio of J/$\psi$ to Drell-Yan production and
in the overall J/$\psi$ yield (see Fig.~\ref{J/psi}). Such an onset
cannot be obtained in any thermal production model, nor can the
centrality-dependent $\psi^\prime$/$\psi$ ratio found in S-U
collisions~\cite{NA38-psi'}. Hence, the present data indicate that the
production and subsequent fate of charmonium states in nuclear
collisions differ significantly from that of light hadrons. We shall
return shortly to the interpretation of the observed charmonium
patterns.

\begin{figure}[ht!]
\vglue-4mm
\centering
\begin{tabular}{ccccc}
\resizebox{0.32\textwidth}{!}{%
\includegraphics*{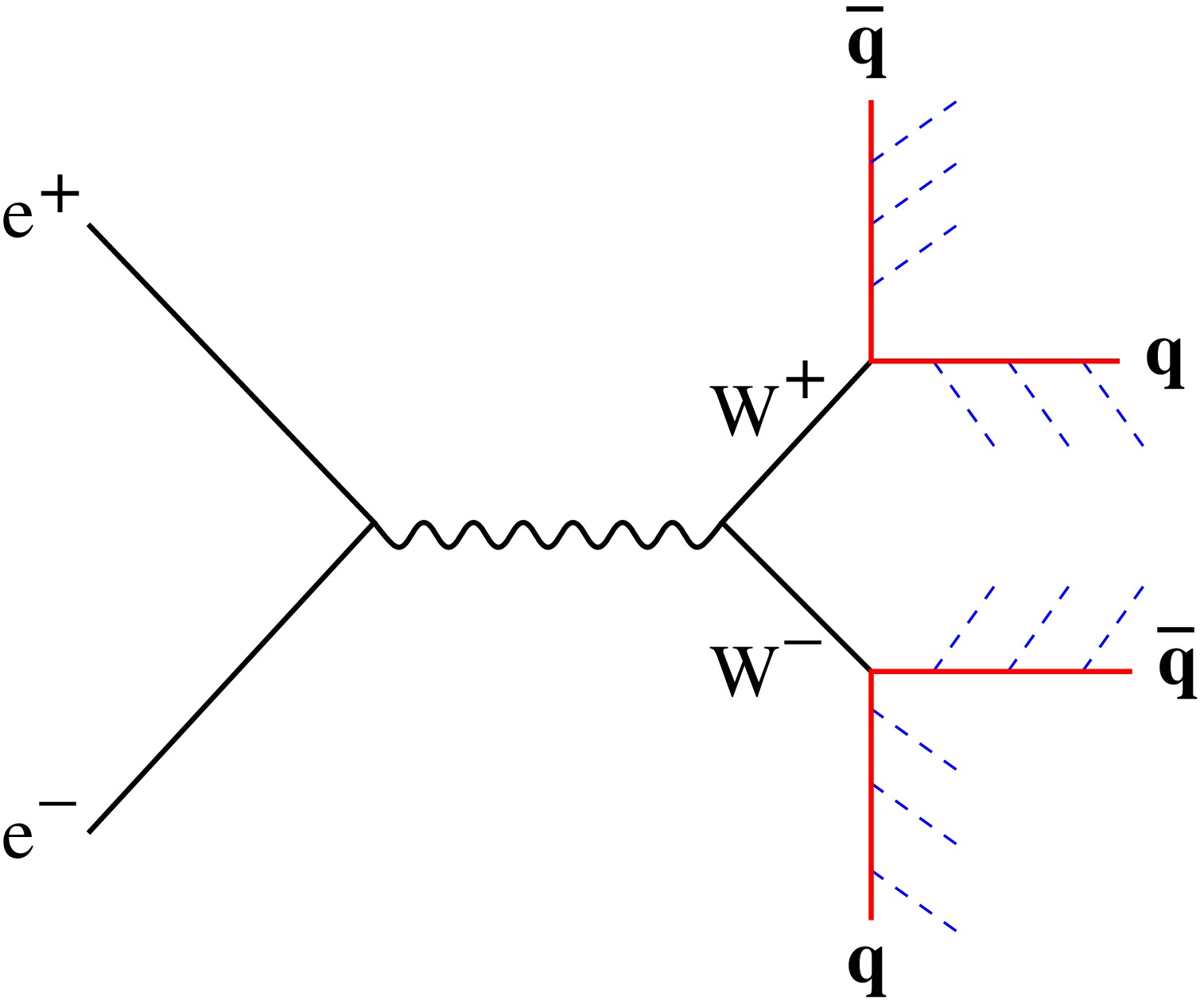}}
& & & &
\resizebox{0.32\textwidth}{!}{%
\includegraphics*{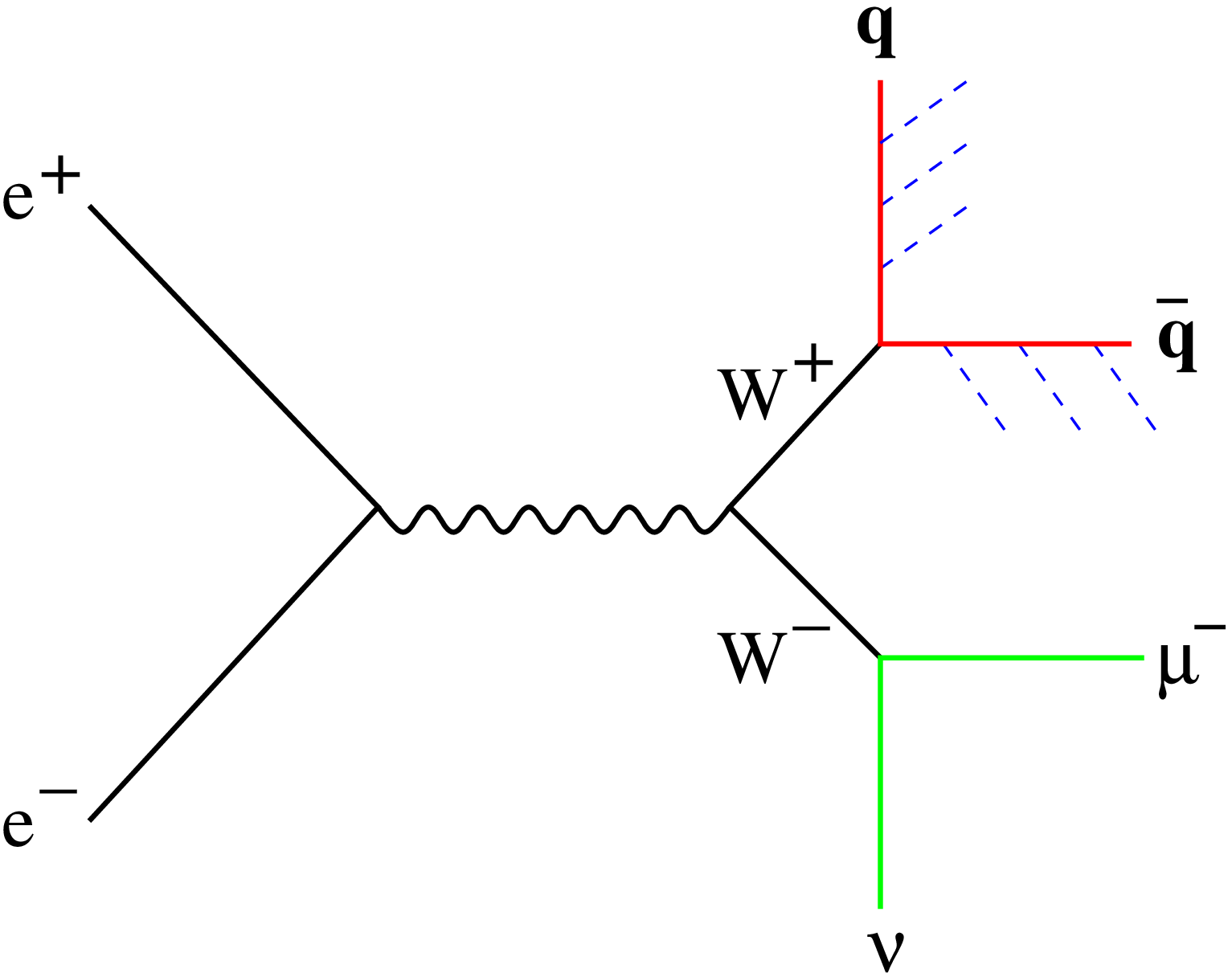}}
\end{tabular}
\vglue-8mm
\caption{Four-jet and two-jet decays of $W^+W^-$ pairs in $e^+e^-$
annihilation at LEP.}
\label{WW}
\vglue-4mm
\end{figure}

The study of the behavior of (light quark) hadrons has thus provided
us with clear indications determining where the hadronic regime
ends. It does not tell us, however, where the QGP begins. High energy
elementary particle collisions lead to initial energy density
estimates higher than those of Pb-Pb collisions at SPS or RHIC
energies (see Fig.~\ref{kabana}); for the resulting hadronization
temperatures we always get the universal $T_H$.  Therefore, one has
tried to find evidence for pre-hadronic interactions, between partons
from different sources. In $e^+e^-$ interactions at LEP energies
($\sqrt s \simeq 200$~GeV), this can be done by comparing hadron
multiplicities in two-jet and four-jet decays of $W^+ W^-$ pairs
produced essentially at rest (see Fig.~\ref{WW}).  `Cross talk'
between jets from different $W$'s should be reflected in hadron
multiplicities, with the prediction \cite{KG} that $N_h(4jet) <
2N_h(2jet)$.  The data do not show any evidence for such `color
interconnection'~\cite{Lepdata}, even though the energy density
calculated according to the Bjorken estimate (\ref{2.1}) reaches
values around 2~GeV/fm$^3$.  The partons from different sources simply
don't seem to know of each other. Color interconnection is clearly a
prerequisite for thermalization on a partonic level; hence $\e \geq
1$~GeV/fm$^3$ is a necessary, but not a sufficient condition for QGP
formation. We must somehow assure that the partons in the early stages
of the collision reach a density so high that they interact and allow
color interconnection to set in. In other words, the pertinent
question is: when do the individual partons from the different
nucleons of the colliding nuclei condense to form a large connected
cluster?

\begin{figure}[b!]
\vglue-4mm
\centering
\begin{tabular}{cccccc}
\resizebox{0.25\textwidth}{!}{%
\includegraphics*{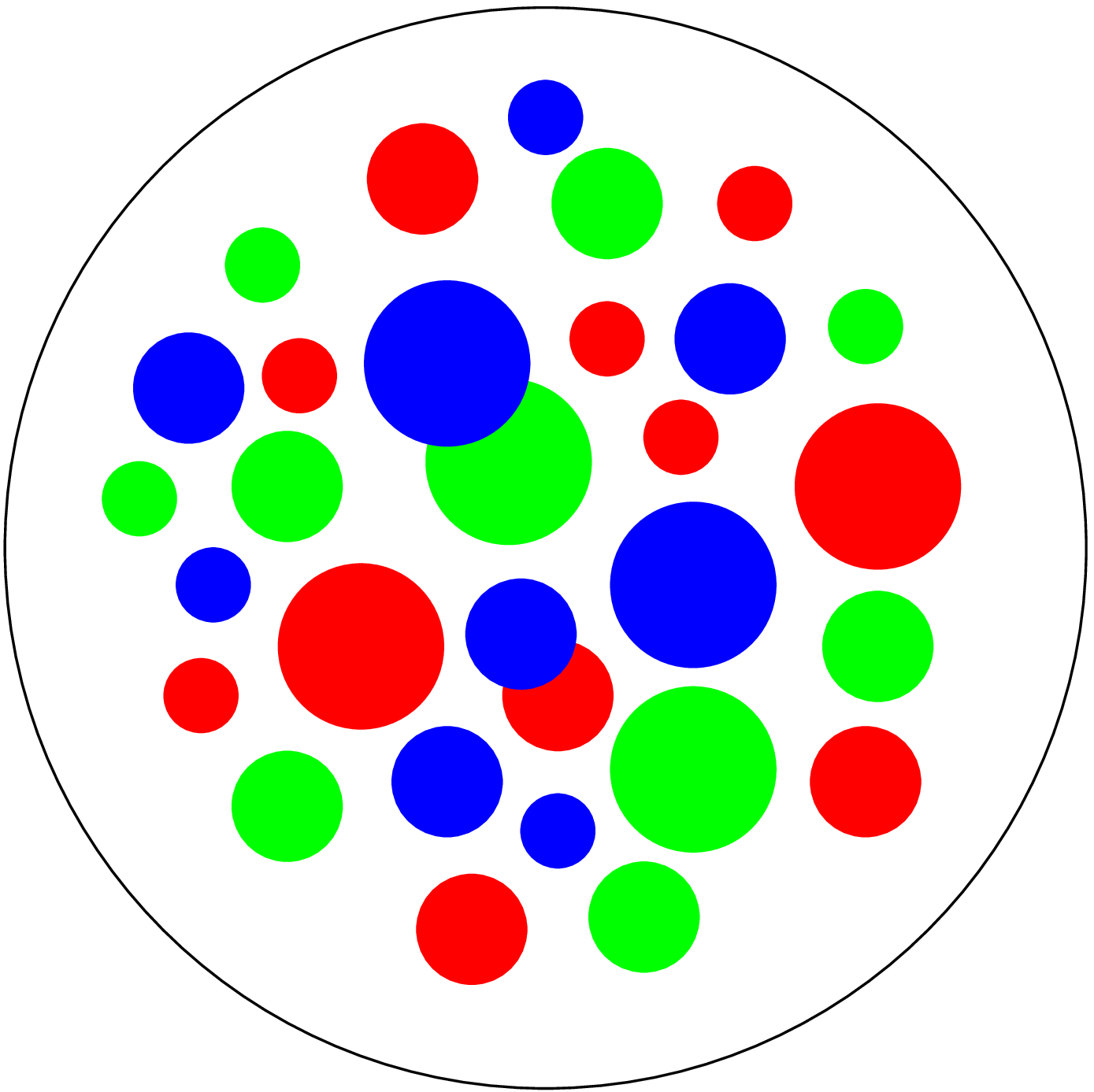}}
& & & & &
\resizebox{0.25\textwidth}{!}{%
\includegraphics*{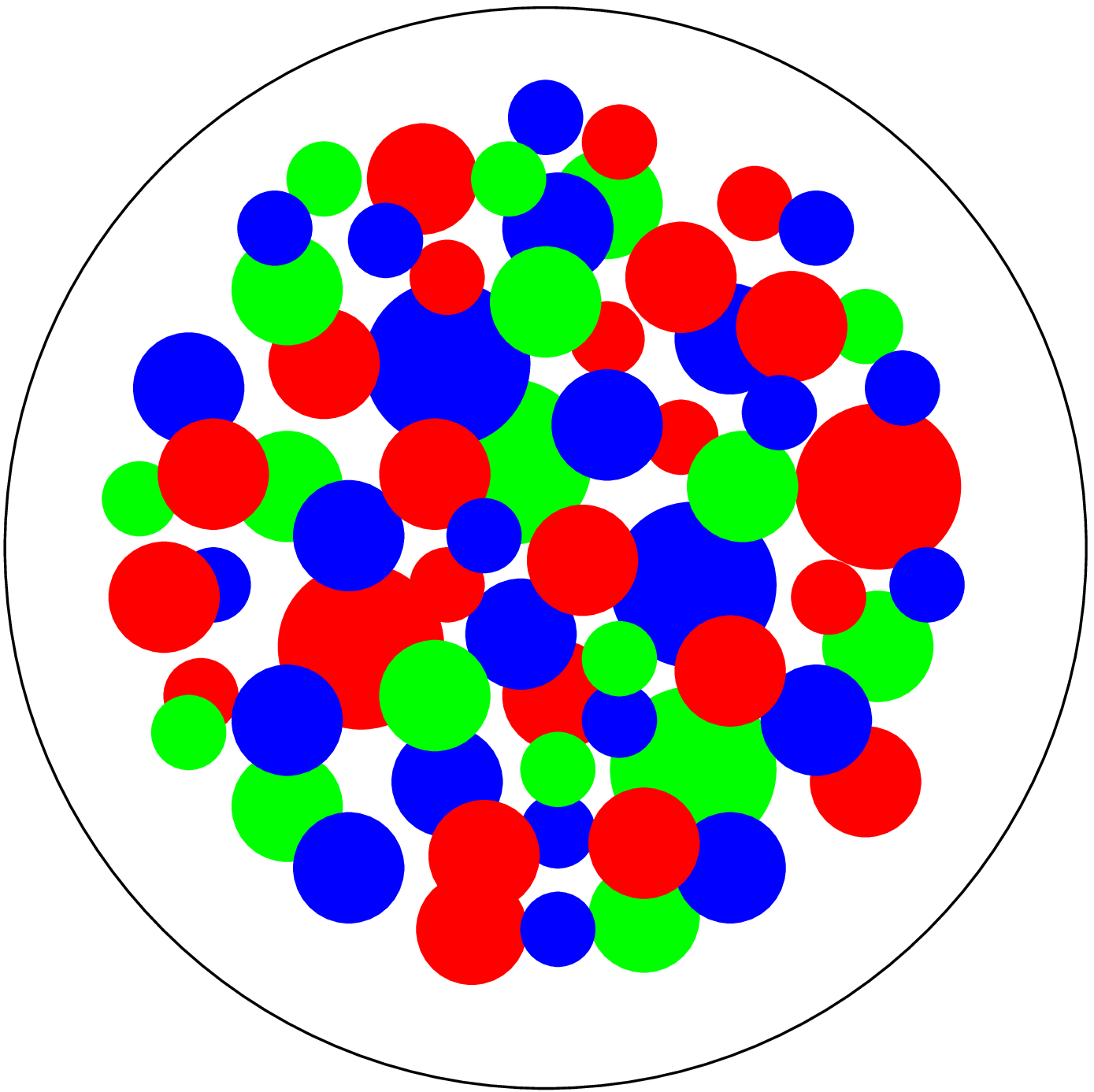}}
\end{tabular}
\vglue-8mm
\caption{Parton distribution in the transverse plane of a
nucleus-nucleus collision, for small (left) and large (right) nuclei.}
\label{PP}
\vglue-4mm
\end{figure}

From percolation theory~\cite{Stauffer}, it is known that the
formation of large-scale clusters is a critical phenomenon; it does
not occur gradually as function of parton density. In the
`thermodynamic' limit of large nuclei, the cluster size diverges at a
specific critical parton density, and even for finite systems it
suddenly increases in a very narrow band of density. Thus there exists
a statistical approach to critical behavior which does not pre-suppose
thermalization and which can be applied to the pre-equilibrium stage
of the nuclear collision evolution \cite{Torino,DFPS}.

In a central high energy nucleus-nucleus collision, the partons in the
two nuclei lead in the transverse plane to an initial condition
schematically illustrated in Fig. \ref{PP}. The transverse size of the
partons is essentially determined by the intrinsic transverse
momentum, $\sigma_q \sim \pi/k_T^2$. The number of partons contained
in a nucleon is known from deep inelastic scattering experiments. It
is parametrized by a parton distribution function depending on the
fraction $x=k/p$ (denoting parton and nucleon momenta by $k$ and $p$,
respectively) and on the scale $Q$ used to resolve the nucleonic
parton structure.  While in lepton-hadron scattering the scale is set
by the virtual photon, in minimum bias nucleon-nucleon or $AA$
collisions it is determined by the transverse momentum of the partons
themselves.

\begin{figure}[ht]
\vglue-6mm
\centering
\resizebox{0.5\textwidth}{!}{%
\includegraphics*{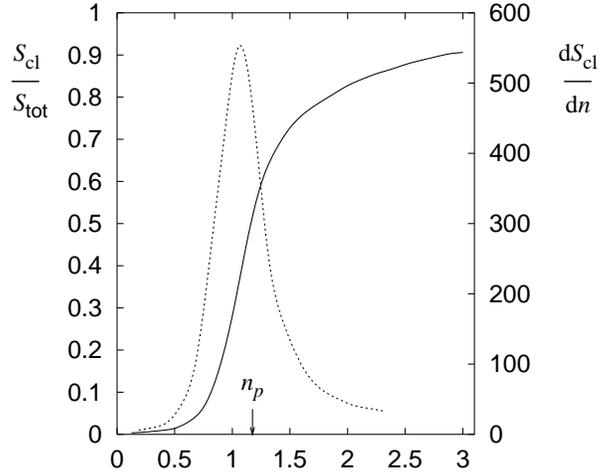}}
\vglue-8mm
\caption{Average cluster size $S_{cl}$ as function of parton density
$n$; $n_p$ denotes the percolation point in the limit $R \to \infty$.}
\label{cluster}
\vglue-6mm
\end{figure}

We denote the nuclear radius in Fig.~\ref{PP} by $R$, the average
parton radius by $r$, and then study the variation of the average
cluster size as function of the parton density $n = N_{\rm parton}/
\pi R^2$.  In the limit $R \to \infty$, the cluster size diverges at
the percolation threshold $n_p \simeq 1.13/ \pi r^2$: 
\be 
S_{cl} \sim (n_p -n)^{-\gamma},
\label{pp1}
\ee
with the critical exponent $\gamma = 43/18$~\cite{isi}. Percolation
thus specifies the onset of connection as a critical phenomenon.  For
finite $R$, there is a pronounced but finite peak at a slightly
shifted density, as shown in Fig.~\ref{cluster}.

To apply this formalism to nuclear collisions~\cite{DFPS}, we need the
transverse parton size and the effective number of partons for a given
collision configuration. The distribution of partons in a nucleon is
determined in the analysis of deep inelastic scattering data; per
nucleon in a nucleon-nucleon collision, at mid-rapidity and for
c.m.s.\ energy $\sqrt s$, one has
\be
N_{\rm parton}(x,Q^2) = \left({dN_q \over dy}\right)_{y=0} 
= xg(x,Q^2) + \sum [xq(x,Q^2) + x{\bar q}(x,Q^2)],
\label{pp2} 
\ee
where $g, q, \bar q$ label the gluon, quark and antiquark
distributions, respectively, and the sum runs over the quark
species. At $y=0$, the fractional parton momentum $x=k/p$ becomes
$x=Q/\sqrt s$. As shown in Fig.~\ref{PP}, there is a distribution of
partons of different transverse sizes up to the resolution scale
$Q^2$; we approximate this by the average value $\pi <\!r^2\!> \simeq
\pi / Q^2$.

In nucleus-nucleus collisions at SPS energies, it is a good
approximation to consider that the activated partons originate from
wounded or participant nucleons.  At higher energies, there will also
exist collision-dependent contributions~\cite{KN}.

We thus obtain for central $AA$ collisions the percolation condition
\be
{2A \over \pi R_A^2} N_{\rm parton}(x,Q^2) = {1.13 \over \pi Q^{-2}};
\label{pp3}
\ee 
it determines the onset of color connection as function of $A$ and of
the c.m.\ collision energy $\sqrt s$. At $\sqrt s = 20$~GeV, we
obtain~\cite{DFPS}:
\smallskip

\hskip 3cm for $A ~\lsim ~60$, no color connection,

\hskip 3cm for $A ~\gsim ~60$, a connected parton condensate.

\smallskip
\noindent
The parton condensate formed beyond the percolation point consists of
overlapping and hence interacting partons from all involved nucleons.
It thus constitutes a deconfined pre-thermal medium, which is a
necessary precursor for QGP formation, since color connection is a
prerequisite for thermalization. The parton condensate is
characterized by a scale $Q=Q_s$, where $Q_s(A,\sqrt s)$ specifies the
onset of percolation for central collisions at a given $A$ and $\sqrt
s$. The average transverse momentum of partons in the condensate is
determined by this scale, so that $Q_s$ is in a sense a precursor of
temperature. Increasing $A$ or $\sqrt s$ leads to higher $Q_s$,
indicating something like a ``hotter" medium.

We have here concentrated on the onset of parton condensation as a
critical phenomenon determined by percolation of quarks and gluons as
geometric entities in the transverse plane.  The behavior of the dense
parton condensate in the limit of large $A$ and $\sqrt s$, the color
glass condensate, has been studied intensively over the past years in
terms of classical color fields. It is becoming increasingly clear
from these studies that the pre-equilibrium stage plays a much more
decisive role in nuclear collisions than previously envisioned. In the
last part of this section, we consider one interesting experimental
consequence of the onset of parton condensation.

To study this onset as function of centrality in
a given $A-A$ collision, we have to replace $2A/\pi R_A^2$ in
Eq.~(\ref{pp3}) by the density of wounded nucleons $n_w(b)$ calculated
for a given impact parameter $b$, using the actual nuclear profile
(Woods-Saxon). For Pb-Pb collisions at $\sqrt s = 17$~GeV, this leads
to
\smallskip 

\hskip 3cm for $N_w ~\lsim ~150$, no color connection,

\hskip 3cm for $N_w ~\gsim ~150$, a connected parton condensate.

\smallskip
\noindent
Instead of the impact parameter, we have here used the number of
wounded nucleons to specify the collision centrality, with $N_w \simeq
390$ for $b = 0$. Note that at this threshold point, the Bjorken
estimate for the energy density gives $\sim 2.5$~GeV/fm$^3$.  This
clearly shows that the onset of color connection only occurs at a much
higher energy density than the value given by finite temperature
lattice QCD for a thermalized medium.

As mentioned above, charmonium suppression in nuclear collisons was
one of the proposed signatures for quark-gluon plasma formation. In
the light of present thinking, we have to consider the fate of the
charmonium states already in the parton condensate phase. This can be
addressed in different ways, invoking generalized color
screening~\cite{goncalves}, dipole break-up in a random color
field~\cite{fuji}, etc. We shall here simply consider the intrinsic
scale $Q_i$ of a given charmonium state $i$ (J/$\psi$, $\chi_c$,
$\psi^\prime$) and assume that when it is exceeded by the condensate
scale,
\be
Q_s > Q_i,
\label{pp4}
\ee 
the charmonium state is dissolved. We are thus assuming that $Q_s$
plays in the pre-equilibrium state the role of the critical
temperature or of the screening mass in a thermal medium.

To show the consequences of this assumption, we must first recall that
J/$\psi$ production in hadronic collisions occurs in part through
feed-down from higher excited states: about 60\,\% of the observed
J/$\psi$ in proton-proton collisions are directly produced $(1S)$
states, the remaining 40\,\% coming from $\chi_c$ ($\sim$\,30\,\%) and
$\psi^\prime$ ($\sim$\,10\,\%) decays. The intrinsic scales of these
states are given by their radii,
\be
r_{\rm J/\psi} \simeq (0.9~{\rm GeV})^{-1},~~
r_{\chi} \simeq (0.6~{\rm GeV})^{-1},~~
r_{\psi^\prime} \simeq (0.6~{\rm GeV})^{-1}.
\label{pp5}
\ee

\begin{figure}[ht!]
\vglue-8mm
\centering
\resizebox{0.55\textwidth}{!}{%
\rotatebox{-90}{%
\includegraphics*{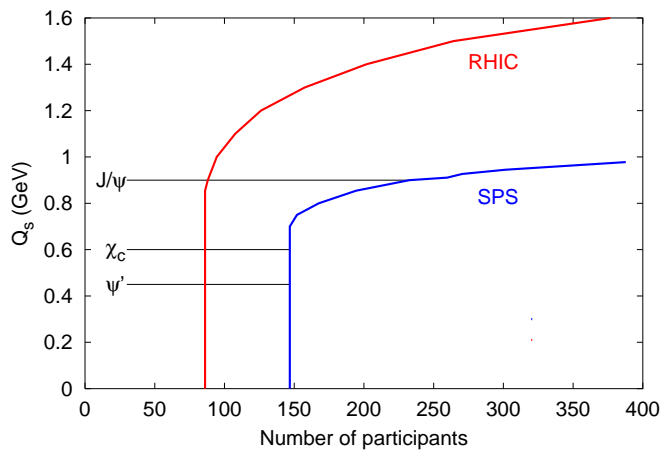}}}
\vglue-8mm
\caption{Centrality dependence of the percolation scale, $Q_s$, in Pb-Pb
collisions, at SPS and RHIC energies~\cite{DFPS}.}
\label{Q-s}
\vglue-4mm
\end{figure}

From the centrality dependence of $Q_s$ shown in Fig.~\ref{Q-s}, we
find that, at SPS energies, charmonia suppression sets in at
\smallskip

\begin{figure}[ht]
\vglue-8mm
\centering
\resizebox{0.55\textwidth}{!}{%
\includegraphics*{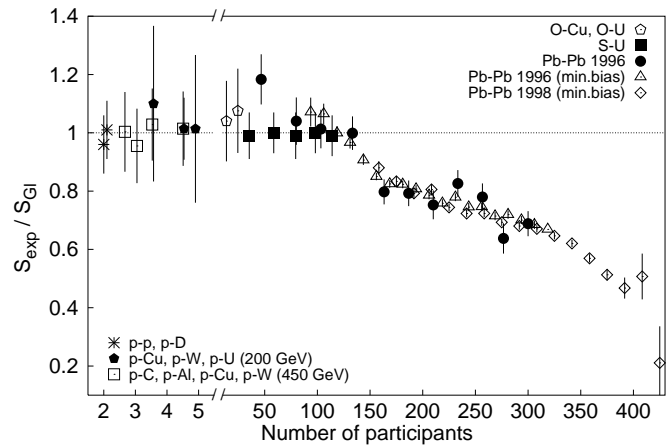}}
\vglue-8mm
\caption{J/$\psi$ production in $pp$, p-A and $AB$ collisions, as 
function of the number of wounded nucleons~\cite{NA50}.}
\label{NA50}
\vglue-4mm
\end{figure}

\hskip 3cm $N_w ~\gsim ~150$ for the $\chi_c$ and $\psi^\prime$ contributions,

\hskip 3cm $N_w ~\gsim ~250$ for direct J/$\psi$ production,

\smallskip
\noindent
to be compared with the observed pattern shown in Fig.~\ref{NA50}.
Obviously these results must be studied in more detail, but it appears
quite likely that the onset of J/$\psi$ suppression indeed indicates
an onset of deconfinement, while not implying any thermalization.

\section{Summary}

So let us summarize what we have learned in the first fifteen years of
ultra-relativistic heavy ion collisions.

We know that there is an intrinsic limit to hadronic matter, in accord
with the phase diagram determined in statistical QCD. The confinement
boundary is established by the SPS/AGS program and agrees, both
qualitatively and quantitatively, with the predictions of lattice QCD.
I believe that the RHIC and LHC experiments can only reconfirm
this. For small $\mu_B$, we know the onset of hadronization, and we
know that strangeness enhancement, transverse momentum broadening and
HBT radii saturate at this onset.

The early stage of the medium produced in nuclear collisions at the
SPS is partonic. With increasing $A$, the partons begin to form
connected clusters, and at a certain critical density, they form a
color condensate consisting of interacting partons from many different
nucleons.  Hence, at this point, color deconfinement begins: partons
no longer have a clear origin or exist in well-defined numbers. The
parton condensate is a necessary precursor of the QGP: it provides the
interacting deconfined partons which, if given enough time for
equilibration, can make a QGP.

Charmonium states of different binding energies and spatial sizes
probe different parton scales of the produced initial state. The
observed step-wise form of anomalous J/$\psi$ suppression appears to
provide the first signal for color deconfinement through parton
condensation; this does not require any thermalization. The
forthcoming J/$\psi$ production measurements at full SPS energy but
with lower $A$, to be performed by NA60, should further clarify this
point. In particular, the comparison between Pb-Pb and In-In J/$\psi$
suppression patterns should determine the critical densities and
scales.

Pioneering results in physics always need to be reconfirmed. But if
these limits of confinement survive the test of time, we will remember
that they were first found in SPS/AGS experiments.

So, at the beginning of the collision evolution at the SPS, there
appears to be color deconfinement; at the end, thermalization and
collective behavior. To produce a quark-gluon plasma medium, we need
to have both at the early stage. Let us see if the future data from
experiments at the much higher RHIC and LHC energies can achieve
this. But no matter how well-defined the road for further exploration
may seem to be, I am sure that the forthcoming studies will rediscover
the one feature which has made this field so challenging and
exciting. It is perhaps best summarized by the Spanish poet Antonio
Machado~\cite{machado}:

\centerline{\sl Caminante, son tus huellas el camino, y nada m\'as;}

\centerline{\sl caminante, no hay camino, se hace camino al andar.}

\smallskip

\centerline{\sl Traveller, the road is nothing more than your footprints;}

\centerline{\sl traveller, there is no road, you make it as you go.}

\section*{Acknowledgements}
It is a pleasure to acknowledge the help of many colleagues in the
preparation of this survey; special thanks go to P.~Braun-Munzinger, 
S.~Digal, S.~Fortunato, F.~Karsch, D.~Kharzeev, M.~Nardi, P.~Petreczky, 
K.\ Redlich, H.\ Specht, U.~Wiedemann and, in particular, to 
C.\ Louren{\c c}o. I~am grateful to L.~V\'azquez for literary support.

\end{document}